\documentclass[floatfix,showpacs,superscriptaddress,prl,twocolumn]{revtex4}
\usepackage{amsbsy,amssymb,amsmath,bm}
\usepackage{graphicx,color}
\usepackage{amsfonts}
\usepackage{amssymb}
\usepackage{amsmath}

\input{tcilatex}

\begin{document}

\title{Signature of Schwinger's pair creation rate via radiation generated
in graphene by strong electric current}
\author{M. Lewkowicz}
\affiliation{\textit{Physics Department, Ariel University Center, Ariel 40700, Israel}}
\author{H.C. Kao}
\affiliation{\textit{Physics Department, National Taiwan Normal University, Taipei 11677,
Taiwan, R. O. C}.}
\author{B. Rosenstein }
\email{vortexbar@yahoo.com}
\affiliation{\textit{Electrophysics Department, National Chiao Tung University, Hsinchu
30050,} \textit{Taiwan, R. O. C}.}
\affiliation{\textit{National Center for Theoretical Sciences, Hsinchu 30043,} \textit{%
Taiwan, R. O. C}.}
\affiliation{\textit{Physics Department, Ariel University Center, Ariel 40700, Israel}}
\date{\today }

\begin{abstract}
Electron - hole pairs are copuously created by an applied electric field
near the Dirac point in graphene or similar 2D electronic systems. It was
shown recently that for sufficiently large electric fields $E$ and ballistic
times the I-V characteristics become strongly nonlinear due to Schwinger's
pair creation rate, proportional to $E^{3/2}$. Since there is no energy gap
the radiation from the pairs' annihilation is enhanced. The spectrum of
radiation is calculated and exhibits a maximum at $\omega =\sqrt{%
eEv_{g}/\hbar }$. \ The angular and polarization dependence of the emitted
photons with respect to the graphene sheet is quite distinctive. For very
large currents the recombination rate becomes so large that it leads to the
second Ohmic regime due to radiation friction.
\end{abstract}

\pacs{72.80.Vp  \  73.20.Mf \ \ 12.20.-m }
\maketitle

\section{Introduction}

Electronic mobility in graphene, especially one suspended on leads, is
extremely large \cite{GeimPRL08} so that a graphene sheet is one of the
purest electronic systems. The relaxation time of charge carriers due to
scattering off impurities, phonons, ripplons, etc., in suspended graphene
samples of submicron length is so large that the transport is ballistic \cite%
{AndreiNN08,BolotinPRL08}. The ballistic flight time in these samples can be
estimated as $t_{bal}=L/v_{g}\,,$ where $v_{g}\simeq 10^{6}m/s$ is the
graphene velocity characterizing the massless "ultra - relativistic"
spectrum of graphene near Dirac points, $\varepsilon _{k}=v_{g}\left\vert 
\mathbf{k}\right\vert $, and $L$ is the length of the sample that can exceed
several $\mu m$ \cite{Andrei10,Vandecasteele10}. The extraordinary physics
appears right at the Dirac point at which the density of states vanishes. In
particular, at this point graphene exhibits a quasi - Ohmic behaviour, $%
\mathbf{J}=\sigma \mathbf{E}$, even in the purely ballistic regime.

A physical picture of this "resistivity" without either charge carriers or\
dissipation is as follows \cite{Sachdev08}. The electric field creates
electron - hole excitations in the vicinity of the Dirac points similar to
the Landau-Zener tunneling effect in narrow gap semiconductors or electron -
positron pair creation in Quantum Electrodynamics first studied by Schwinger 
\cite{Schwinger} (later referred to as LZS). Importantly, in graphene the
energy gap is zero, thus the pair creation is possible at zero temperature
and arbitrary small $\mathbf{E,}$ even within linear response. Although the
absolute value of the quasiparticle velocity $v_{g}$ cannot be altered by
the electric field due to the "ultra - relativistic" dispersion relation%
\textbf{, }the orientation of the velocity can be influenced by the applied
field.\textbf{\ }The electric current, $e\mathbf{v}$, proportional to the
projection of the velocity $\mathbf{v}$ onto the direction of the electric
field is increased by the field. These two sources of current, namely
creation of moving charges by the electric field (polarization) and their
reorientation (acceleration) are responsible for the creation of a stable
current.

Agreement over the qualitative explanation notwithstanding, determination of
the value of the minimal DC conductivity at Dirac point in the limit of zero
temperature had undergone a period of experimental and theoretical
uncertainty. After the value in graphene on substrate was measured to be
about $\sigma =4e^{2}/h$ \cite{Novoselov05}, it was shown in experiments on
suspended samples \cite{AndreiNN08} that the zero temperature limit was not
achieved and in fact that these early samples had too many charged
"puddles", so that they represented an average around the neutrality or the
Dirac point. The value in early-on suspended samples \cite{AndreiNN08} was
half of that and most recently settled at the "dynamical" $\sigma _{2}=\frac{%
\pi }{2}\frac{e^{2}}{h}$ in best samples at $2K$ temperature \cite{Andrei10}%
. Theoretically several different values appeared. The value $\sigma _{1}=%
\frac{4}{\pi }\frac{e^{2}}{h}$ had been considered as the "standard" one for
several years \cite{Castro} and appeared as a zero disorder limit in many
calculations like the self consistent harmonic approximation, although
different regularizations within the Kubo formalism resulted in different
values \cite{Ziegler06}.

The dynamical approach to transport was applied to the tight binding model
of graphene \cite{Lewkowicz09} to resolve this "regularization ambiguity".
It consists of considering the ballistic evolution of the current density in
time after a sudden or gradual switching on of the electric field. The
result within linear response is that the current settles very fast, on the
microscopic time scale of $t_{\gamma }=\hbar /\gamma \simeq 0.24$ $fs$ ($%
\gamma $ being the hopping energy), on the value of $J=\sigma _{2}E$. The
value is identical to the one obtained (at nonzero temperatures) for the AC
conductivity\cite{Varlamov07}. The two contributions, polarization and
attenuation are comparable in strength and combine to produce a constant
total current. However a deeper analysis of the "quasi - Ohmic" graphene
system beyond the leading order in perturbation theory in electric field
revealed \cite{Rosenstein10} that on the time scale 
\begin{equation}
t_{nl}=\sqrt{\frac{\hbar }{eEv_{g}}},  \label{t_nl}
\end{equation}%
the linear response breaks down. For larger times the quasi - Ohmic behavior
no longer holds. This is in contrast to dissipative systems, in which the
linear response limit can be taken directly at infinite time. This perhaps
is the origin of the "regularization" ambiguities in graphene, since large
time and small field limits are different. The time scale on which nonlinear
effects become dominant is not always very large; for example, in
experiments dedicated to breakdown of Quantum Hall effect \cite{SinghPRB09}
in which $E=10^{4}V/m$, nonlinearity sets in at $t_{nl}=0.3ps$, that is of
order ballistics time for $L=0.3\mu m$. Graphene flakes under larger fields
of order $2\cdot 10^{6}V/m$ have been studied very recently (at room
temperature) in specially designed high current density experiments \cite%
{Vandecasteele10}. In this case the nonlinear time is only $20fs,$ much
lower than the ballistic time $t_{bal}=2ps$ for $L=2\mu m$.\ Analytic and
numerical solutions of the tight binding model\cite{Rosenstein10}, as well
as of the Dirac model describing the physics near the Dirac point
demonstrated\cite{Rosenstein10,Dora10} that at $t_{nl}$ the electron - hole
pairs creation becomes dominant and is well described by an adaptation of
the well - known (non-analytic in $\mathbf{E}$) Schwinger electron -
positron pair creation rate 
\begin{equation}
\frac{d}{dt}N_{p}=\frac{3^{3/4}}{2^{9/2}v_{g}^{1/2}}\left( \frac{eE}{\hbar }%
\right) ^{3/2}.  \label{Schwinger_rate}
\end{equation}%
The difference with the original derivation \cite{Schwinger} in the context
of particle physics is that the fermions are 2+1 dimensional and "massless",
thus magnifying the effect. The polarization current is $J\left( t\right)
=2ev_{g}N\left( t\right) $ and therefore Schwinger's creation rate leads to
a linear increase with time\cite{Rosenstein10}: 
\begin{equation}
J\left( t\right) =\sigma _{2}\left( \frac{\sqrt{3}}{2}E\right) ^{3/2}\left( 
\frac{ev_{g}}{\hbar }\right) ^{1/2}t\text{.}  \label{J_nl}
\end{equation}

The physics of pair creation is highly non-perturbative and non-linear in
nature and therefore, instead of the linear response, Schwinger found an
exact formula using functional methods. The rate can be intuitively
understood using the much simpler instanton approach originally proposed in
the context of particle physics \cite{Nussinov} (extended later to low
dimensions \cite{Gavrilov96}), but is known in fact in condensed matter
physics as the Landau - Zener tunneling probability \cite%
{Dora10,Santoro,Vandecasteele10}. In particle physics it is extremely
difficult to observe Schwinger's creation rate and it would be interesting
to establish experimentally this dynamical phase in low dimensional
condensed matter physics featuring the massless Dirac quasiparticle spectrum
like graphene or novel materials sharing with it the massless Dirac spectrum
like topological insulators or tuned semiconductor heterojunctions \cite%
{topins}. Of course transport phenomena at rather large fields always have a
background related to possible influence of leads, disorder and thermal
effects like local heating, etc.

In this note we draw attention to a direct and unintrusive signature of the
dynamical phase of LZS pair creation in a graphene sheet subject to an
applied electric field. It is demonstrated that the flux of photons radiated
from the surface of the sample is characterized by the creation rate since
the photons are emitted via electron - hole pair annihilation and therefore
proportional to $E^{3/2}$, a hallmark of Schwinger's process. In addition,
the frequency, direction and polarization characteristics of the radiation
generated by the electric field calculated here all bear footprints\ of the
pair creation dynamics.

\section{Electron - hole recombination rate into photons.}

\subsection{Amplitude for emission of a single photon}

The electrons and their electromagnetic interaction with photons are
approximately described near a Dirac point by the Weyl Hamiltonian: 
\begin{equation}
H=\int d^{3}r\text{ }\psi ^{\dagger }\left[ 
\begin{array}{c}
v_{g}\mathbf{\sigma }\cdot \left( -i\hbar \mathbf{\nabla }+\frac{e}{c}%
\mathbf{A}\right) \\ 
-\frac{\hbar ^{2}}{2m}\left( \partial _{z}+i\frac{e}{\hbar c}A_{z}\right)
^{2}+V_{conf}\left( z\right)%
\end{array}%
\right] \psi .  \label{H}
\end{equation}%
Here $\psi $ is the two component spinor second quantized field and $\left( 
\mathbf{A,}A_{z}\right) $ is the vector potential (bold letters describe
vectors in the graphene plane, while $z$ is the perpendicular direction).
Electrons (charge $-e$) and holes (charge $e$) in the graphene sheet are
confined in this model to the $z=0$ plane by a potential $V_{conf}$ (small
shape changes can be neglected for our purposes). The only requirement from
this potential is that it is strong enough to "freeze" the motion along the $%
z$ direction. In the single graphene sheet one has two left handed chirality
Weyl fermions described by the above Hamiltonian in which $\mathbf{\sigma }$
denotes the in - plane Pauli matrices and two right handed Weyl fermions
represented by $\mathbf{\sigma }^{\dagger }$. To include the topological
insulators case \cite{topins}, we first concern ourselves with only one
spinor.


\begin{center}
\includegraphics[
width=.5\columnwidth]
{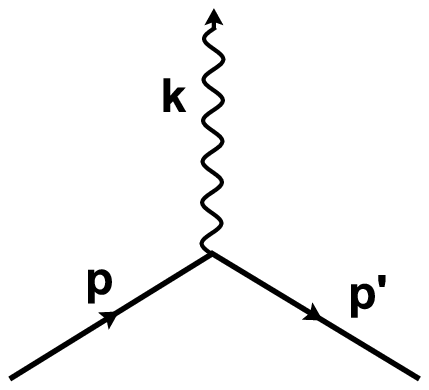}\\
Fig.1. Feynman diagrams representing the major electromagnetic processes in graphene. a. The one photon emission.
\label{Fig.1a}%
\end{center}

\begin{center}
\includegraphics[
width=.7\columnwidth]
{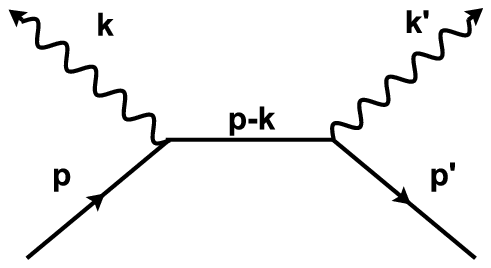}\\
b. The two photon emission
\label{Fig. 1b}%
\end{center}
\bigskip

We consider the emission of a photon with wave vector $\left( \mathbf{k,}%
k_{z}\right) $ and frequency $\omega =c\sqrt{\mathbf{k}^{2}+k_{z}^{2}}$,
described by a linearly polarized plane wave, 
\begin{equation}
\mathbf{A}_{ph}\mathbf{=}\frac{2E_{0}}{\omega }\mathbf{e}^{\left( \lambda
\right) }\sin \left( \mathbf{k\cdot r+}k_{z}z-\omega t\right) ,  \label{Aext}
\end{equation}%
whereas the DC applied field is $\mathbf{A}_{ext}\mathbf{=}\left(
0,-cEt\right) $. For regularization we make use of a finite box $L\times
L\times L_{z}$, so that momenta are discrete and the single photon's
electric field is $E_{0}^{2}=\hbar \omega /\left( L^{2}L_{z}\right) $. The
unit vectors 
\begin{eqnarray}
\mathbf{e}^{\left( 1\right) } &=&\left( -\sin \varphi ,\cos \varphi \right) ;%
\text{ }e_{z}^{\left( 1\right) }=0;  \label{e_lambda} \\
\mathbf{e}^{\left( 2\right) } &=&-\cos \theta \left( \cos \varphi ,\sin
\varphi \right) ;\text{ }e_{z}^{\left( 2\right) }=\sin \theta ,  \notag
\end{eqnarray}%
describe polarizations that are conveniently chosen similarly to a recent
calculation of electromagnetic emission due to thermal fluctuations \cite%
{thermal}. The vectors $\mathbf{e}^{\left( 1\right) }$ and $\mathbf{e}%
^{\left( 2\right) }$ represent the "in plane" and the "out of plane"
polarizations, respectively. The electron and the hole wave functions are $%
\frac{1}{\sqrt{2}L}e^{i\mathbf{p\cdot r}}u\left( \mathbf{p}\right) \psi
_{n}\left( z\right) $ and $\frac{1}{\sqrt{2}L}e^{i\mathbf{p}^{\prime }\cdot 
\mathbf{r}}v\left( \mathbf{p}^{\prime }\right) \psi _{n}\left( z^{\prime
}\right) $, with spinors defined by%
\begin{equation}
u\left( \mathbf{p}\right) =\left( 
\begin{array}{c}
1 \\ 
-ie^{i\phi }%
\end{array}%
\right) ;\text{ }v\left( \mathbf{p}^{\prime }\right) =\left( 
\begin{array}{c}
1 \\ 
ie^{i\phi ^{\prime }}%
\end{array}%
\right) ,  \label{u}
\end{equation}%
with $\mathbf{p+}\frac{e}{\hbar c}\mathbf{A}_{ext}\mathbf{=}p\left( \cos
\phi ,\sin \phi \right) .$ $\psi _{n}\left( z\right) $ are wave functions of
the confinement. The interaction with a photon at time $t$ happens when the
momentum is minimally shifted due to the DC field. The Golden rule photon
emission rate (for an "initial" electron with momentum $\mathbf{p}$ and \ a
"final" hole $\mathbf{p}^{\prime }$ and a photon of polarization $\lambda $
and momentum $\left( \mathbf{k},k_{z}\right) $) is

\begin{eqnarray}
W_{nn^{\prime }}^{\left( \lambda \right) }\left( \mathbf{p},\mathbf{p}%
^{\prime },\mathbf{k},k_{z},t\right) &=&\frac{2\pi }{\hbar }\left\vert
F_{nn^{\prime }}^{\left( \lambda \right) }\right\vert ^{2}N_{\mathbf{p}%
}\left( t\right) N_{-\mathbf{p}^{\prime }}\left( t\right)  \label{W_lambda}
\\
&&\times \delta \left( \hbar v_{g}\left( p+p^{\prime }\right) -\hbar \omega
\right) \text{.}  \notag
\end{eqnarray}%
In terms of Feynman diagrams of Quantum Electrodynamics\cite{Ahiezer} it
corresponds to the diagram in Fig. 1a. Here\ $N_{\mathbf{p}}\left( t\right) $
is the density of electrons in a certain momentum range produced by the
electric field $\mathbf{E}$, and $N_{-\mathbf{p}^{\prime }}\left( t\right) $
the density of holes (equal to that of the electrons at the opposite
momentum due to particle - hole symmetry). The density calculated using the
simple Landau - Zener creation rate expression for one of the flavours is 
\cite{Nussinov,Rosenstein10,Dora10}: 
\begin{equation}
N_{\mathbf{p}}\left( t\right) =\Theta \left( p_{y}\right) \Theta \left( 
\frac{e}{\hbar }Et-p_{y}\right) \exp \left( -\frac{\pi \hbar v_{g}}{eE}%
p_{x}^{2}\right) \text{,}  \label{LZ}
\end{equation}%
where $\Theta $ are the Heaviside functions. The transition amplitude is
given by

\begin{eqnarray}
F_{nn^{\prime }}^{\left( \lambda \right) } &=&i\frac{E_{0}}{\omega }\frac{%
ev_{g}}{2L^{2}}e^{i\left( v_{g}\left( p+p^{\prime }\right) -\omega \right) }%
\mathcal{F}_{\mathbf{p},\mathbf{p}^{\prime }}^{\left( \lambda \right) }
\label{M} \\
&&\int dz\text{ }e^{ik_{z}z}\psi _{n}^{\ast }\left( z\right) \psi
_{n^{\prime }}\left( z\right) \delta \left( \mathbf{p+p}^{\prime }-\mathbf{k}%
\right) ,  \notag
\end{eqnarray}%
where matrix elements $\mathcal{F}_{\mathbf{p},\mathbf{p}^{\prime }}^{\left(
\lambda \right) }\equiv v^{\dagger }\left( -\mathbf{p}^{\prime }\right) 
\mathbf{\sigma }\cdot \mathbf{e}^{\left( \lambda \right) }u\left( \mathbf{p}%
\right) $ are%
\begin{eqnarray}
\left\vert \mathcal{F}_{\mathbf{p},\mathbf{p}^{\prime }}^{\left( 1\right)
}\right\vert ^{2} &=&2\left[ 1-\cos \left( 2\varphi -\phi -\phi ^{^{\prime
}}\right) \right] ;  \label{matrix_elements} \\
\left\vert \mathcal{F}_{\mathbf{p},\mathbf{p}^{\prime }}^{\left( 2\right)
}\right\vert ^{2} &=&2\cos ^{2}\theta \left[ 1+\cos \left( 2\varphi -\phi
-\phi ^{^{\prime }}\right) \right] \text{.}  \notag
\end{eqnarray}

\subsection{Spectral emittance}

For tight confinement to the $z=0$ plane one should consider only the ground
state $n=n^{\prime }=0$. Note that the perpendicular component of the wave
vector $k_{z}$ is "free" from conservation that prohibits the process in
fully relativistic QED\cite{Ahiezer}. The phase space for annihilation is
very limited due to $v_{g}<<c$, see Appendix A and leads to important
simplifications.

\begin{center}
\includegraphics[
natheight=2.6723in, natwidth=3.9972in, height=2.3964in, width=3.5743in]
{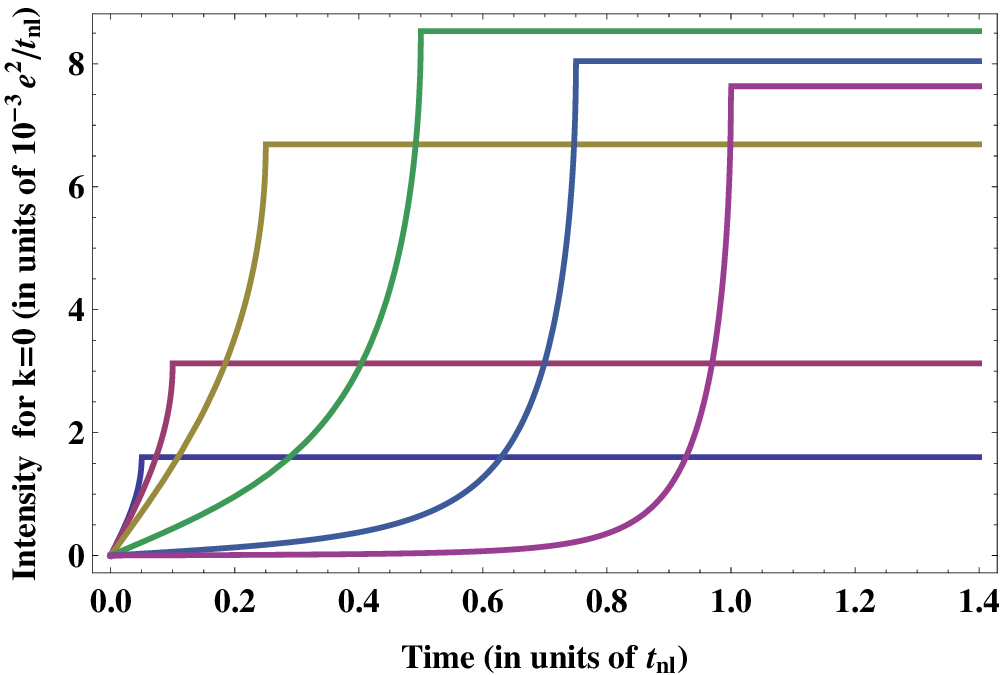}\\
Fig.2. The spectral emittance in direction perpendicular to the graphene plane, 
$\mathbf{k}=0$. Polarizations are summed over. The emittance (in units of $%
e^{2}/t_{nl})$ for various frequencies (in units of $t_{nl}^{-1}$) as
function of ballistic time from $0.1t_{nl}$ to $1.4t_{nl}$.\label{Fig. 2}%
\end{center}

\begin{center}
\includegraphics[
natheight=2.4915in, natwidth=3.9972in, height=2.2027in, width=3.5189in]
{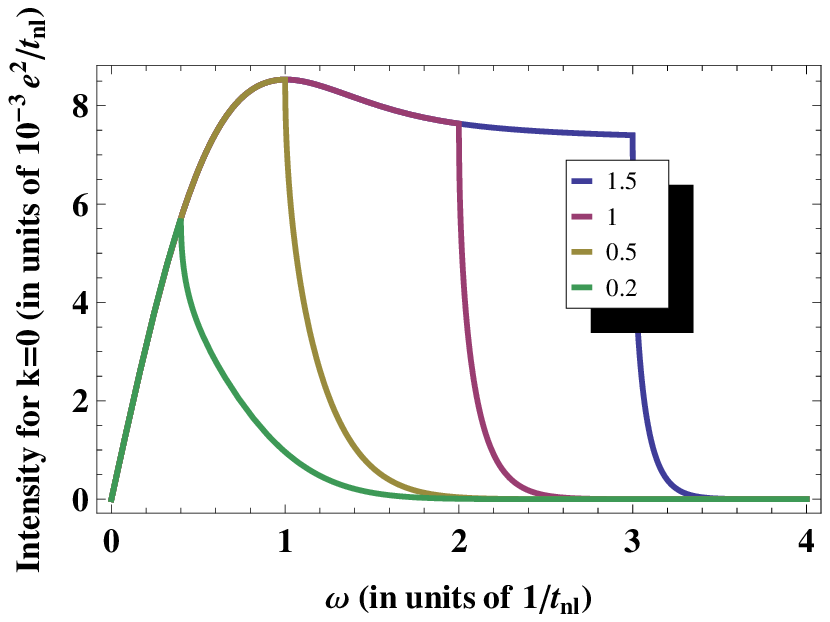}\\
Fig.3. The emittance at various times (in units of $t_{nl}$) as function of
frequency.\label{Fig. 3}%
\end{center}

Let us define the spectral emittance per volume of the $k$ - space (and area
of the graphene flake) as

\begin{eqnarray}
\mathcal{M}^{\left( \lambda \right) }\left( \mathbf{k},k_{z},t\right) &=&%
\frac{4\hbar \omega }{L^{2}}\dsum\nolimits_{\mathbf{p},\mathbf{p}^{\prime }}%
\frac{d}{dk_{z}d\mathbf{k}}W^{\left( \lambda \right) }\left( \mathbf{p},%
\mathbf{p}^{\prime },\mathbf{k},k_{z},t\right)  \notag \\
&=&\frac{e^{2}v_{g}^{2}}{\left( 2\pi \right) ^{4}}\int d\mathbf{p}\left\vert 
\mathcal{F}_{\mathbf{p},\mathbf{k}-\mathbf{p}}^{\left( \lambda \right)
}\right\vert ^{2}N_{\mathbf{p}}N_{\mathbf{k}-\mathbf{p}}  \notag \\
&\times &\delta \left( v_{g}\left( p+\left\vert \mathbf{k-p}\right\vert
\right) -\omega \right) \text{,}  \label{Idef1}
\end{eqnarray}%
where the integration over $\mathbf{p}^{\prime }$ was performed using the
delta function expressing the conservation of momentum. We first study the
frequency dependence of the radiation in the direction perpendicular to the
graphene flake, $\mathbf{k=0}$. Multiplying with $4$ for the spin and valley
degeneracy, summing over the polarizations $\left( \dsum\nolimits_{\lambda
}\left\vert \mathcal{F}_{\mathbf{p},\mathbf{p}}^{\left( \lambda \right)
}\right\vert ^{2}=4\right) $, and integrating over $\mathbf{p}$ one obtains,
using $\omega =ck_{z},$ the spectral emittance%
\begin{eqnarray}
\mathcal{M}\left( \mathbf{k=0,}\omega ,t\right) &=&\frac{%
e^{2}v_{g}^{2}t_{nl}^{2}}{\pi ^{4}}\int_{-t/t_{nl}}^{0}d\overline{p}\text{ }%
\Theta \left( t_{nl}\omega /2+\overline{p}\right)  \notag \\
&&\times \frac{\exp \left[ -2\pi \left( t_{nl}^{2}\omega ^{2}/4-\overline{p}%
^{2}\right) \right] }{\omega \left( t_{nl}^{2}\omega ^{2}/4-\overline{p}%
^{2}\right) ^{1/2}}\text{.}  \label{I_perp1}
\end{eqnarray}

The spectral emittance, presented in Fig.2 for various frequencies as
function of time, increases linearly for $t<<\omega ^{-1}$, $\mathcal{M}%
\left( \mathbf{k=0,}\omega ,t\right) =\frac{2e^{2}t}{\pi ^{4}t_{nl}^{2}}%
e^{-\pi t_{nl}^{2}\omega ^{2}/2}$, then rises sharply approaching a maximum
at $t=\omega t_{nl}^{2}/2$ and stabilizes at 
\begin{equation}
\mathcal{M}\left( \mathbf{k=0,}\omega ,t>>t_{nl}\right) =\frac{e^{2}}{\pi
^{3}}\omega e^{-\pi t_{nl}^{2}\omega ^{2}/4}I_{0}\left( \frac{\pi
t_{nl}^{2}\omega ^{2}}{4}\right) \text{,}  \label{Iasympt}
\end{equation}%
where $I_{0}$ is the modified Bessel function. The asymptotic value rises
linearly with frequency, $\pi ^{-3}\omega $, in the infrared, reaches its
maximum at $\omega =t_{nl}^{-1}$ and falls slightly to $\frac{\sqrt{2}e^{2}}{%
\pi ^{4}t_{nl}}$ in the ultraviolet. In Fig.3 the emittance at various
ballistic times is given as function of frequency. For each ballistic time
the curve has two parts. The first follows the universal dependence given by
Eq.(\ref{Iasympt}). Therefore the frequency for observation of the Schwinger
effect, not surprisingly, should exceed $\omega _{\min }=\sqrt{eEv_{g}/\hbar 
}$, that amounts to $3.6THz$ for $E=10^{4}V/m,$ and $50THz$ for $E=2\cdot
10^{6}V/m$. \ At a higher frequency $\omega _{\max }=2t/$ $t_{nl}^{2}$ the
emittance sharply drops. Therefore the frequency does not exceed $%
2t_{bal}/t_{nl}^{2}$.

\begin{center}
\includegraphics[
natheight=6.9297in, natwidth=11.3455in, height=2.8106in, width=4.5844in]
{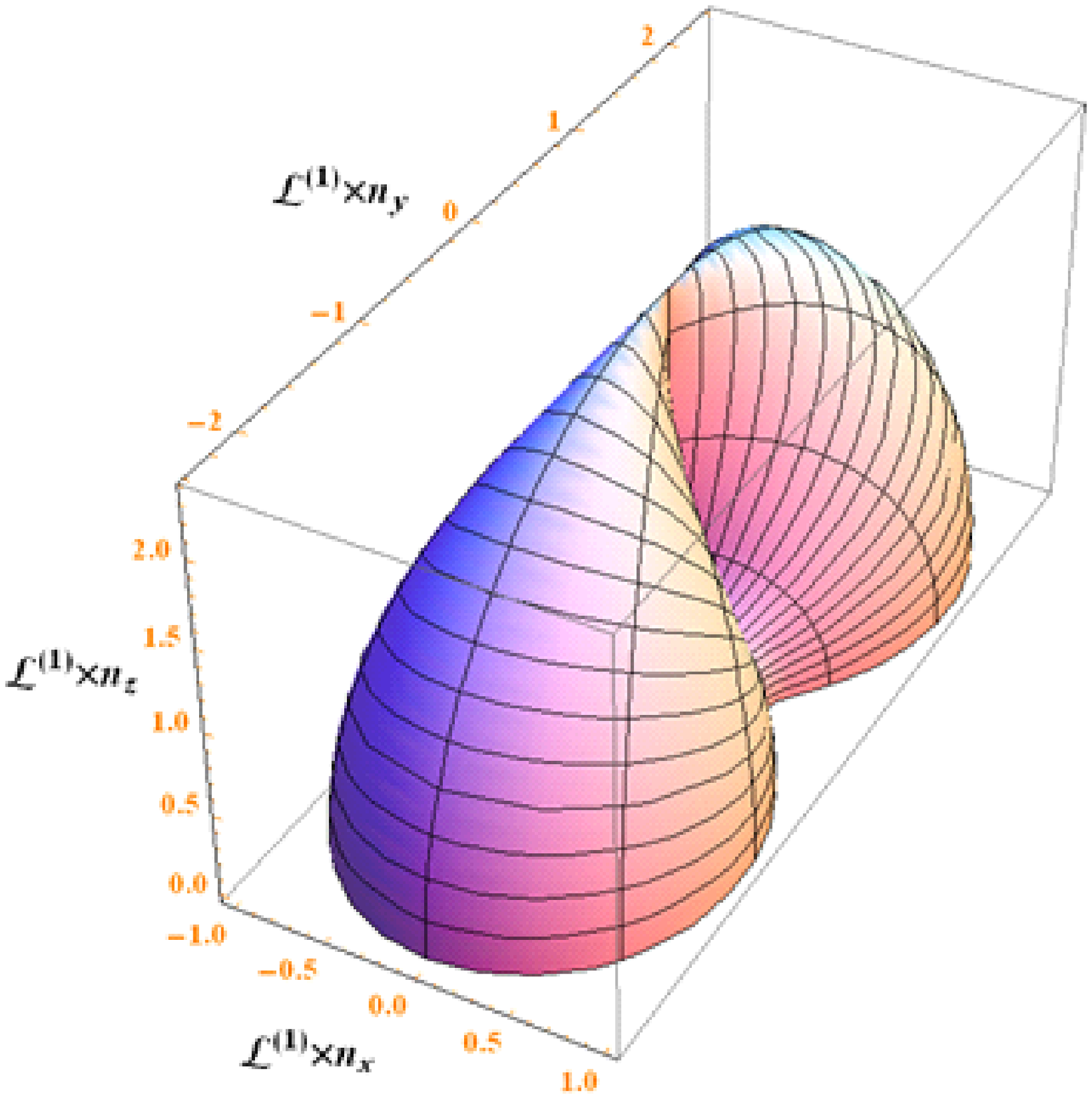}\\
Fig.4a. The angle dependence of the intensity as function of the photon spherical angles 
$\varphi ,\theta $ (half of the whole solid angle). Time is fixed at $%
t=t_{nl}.$ The in - plane polarization. \label{. }%
\end{center}

\begin{center}
\includegraphics[
natheight=9.2206in, natwidth=9.0926in, height=2.7155in, width=2.6783in]
{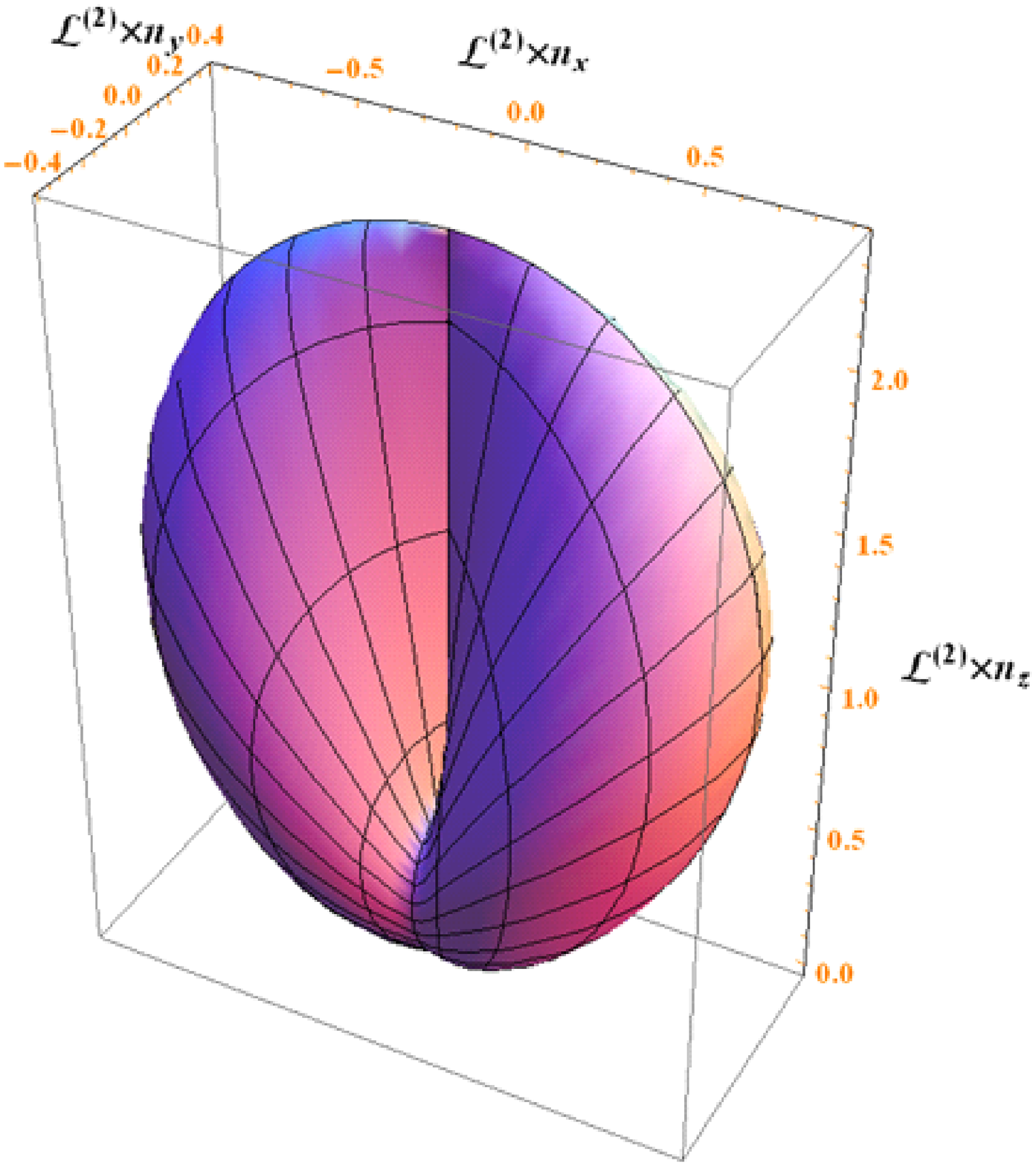}\\
Fig.4b. The out of plane polarization.
\end{center}

\subsection{Angular and polarization distribution}

Next we consider the angular and polarization dependence of the radiated
power per unit area defined as the spectral intensity, Eq(\ref{Idef1}),
integrated over frequencies:%
\begin{equation}
\mathcal{L}^{\left( \lambda \right) }\left( \theta ,\varphi ,t\right) \equiv
\int_{0}^{\infty }d\omega \frac{\omega ^{2}}{c^{3}}\mathcal{M}^{\left(
\lambda \right) }\left( \mathbf{k},\frac{\omega }{c},t\right) .
\label{I_lambda_def}
\end{equation}%
Performing integrations and simplifying, utilizing the small parameter $%
v\equiv v_{g}/c\simeq 1/300<<1$, see Appendix B for details, one obtains:%
\begin{eqnarray}
\mathcal{L}^{\left( 1\right) }\left( \varphi ,t\right) &=&\frac{%
e^{4}v^{4}E^{2}}{2^{5/2}\pi ^{4}c\hbar ^{2}}\left( \frac{t}{4\pi t_{nl}}\cos
^{2}\varphi +\frac{t^{3}}{3t_{nl}^{3}}\sin ^{2}\varphi \right) ;  \notag \\
&&  \label{angle1}
\end{eqnarray}

\begin{eqnarray}
\mathcal{L}^{\left( 2\right) }\left( \theta ,\varphi ,t\right)  &=&\frac{%
e^{4}v^{4}E^{2}}{2^{5/2}\pi ^{4}c\hbar ^{2}}\cos ^{2}\theta   \notag \\
&&\times \left( \frac{t}{4\pi t_{nl}}\sin ^{2}\varphi +\frac{t^{3}}{%
3t_{nl}^{3}}\cos ^{2}\varphi \right) \text{.}  \label{angle21a}
\end{eqnarray}%
The radiant flux from a flake of a $\mu m\times \mu m$ size is $4.7\cdot
10^{-21}W$, for $E=10^{4}V/m$ corresponding to the emission rate of just $10$
photons per second, yet for the high current samples\cite{Vandecasteele10}
with $E=2\cdot 10^{6}V/m$ of the same area one gets a more significant
output: the radiant flux is $1.3\cdot 10^{-17}W$, corresponding to the
emission rate of $3\cdot 10^{4}$ photons per second.

\begin{center}
\includegraphics[
natheight=8.0851in, natwidth=8.0402in, height=3.3425in, width=3.3243in]
{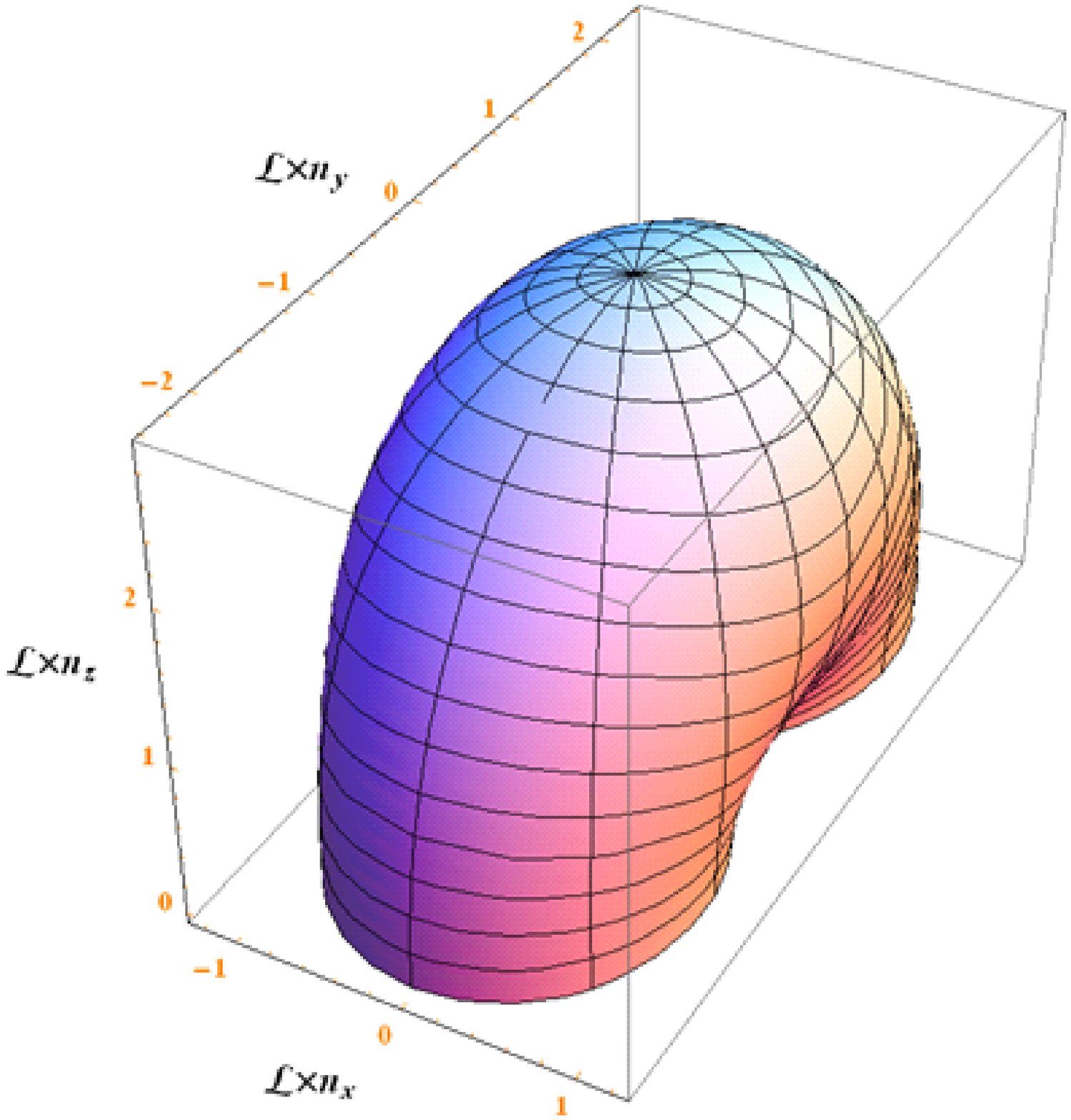}\\
Fig.4c Unpolarized light.
\end{center}

The two quantities $\mathcal{L}^{\left( 1,2\right) }$ and their sum are
presented for $t=t_{nl}$ in the spherical plots Fig.4a-c, respectively. The
radiated power is maximal in direction perpendicular to the graphene plane.
For directions close to the azimuth angles $\varphi =0^{\circ }$ and $%
180^{\circ }$ (perpendicular to the electric field or current) at small
polar angles $\theta $ ( perpendicular to the graphene plane) the
intensities of the two polarizations are of the same order, while for $%
\theta \sim 90^{\circ }$ (close to the in - plane direction) the "out of
plane" polarization, $\lambda =2$, dominates. On the other hand, for $%
0^{\circ }<<\varphi <<180^{\circ }$ the picture is opposite. As expected,
the unpolarized intensity, Fig. 2c, is less anisotropic; yet the radiation
is somewhat depressed in the direction perpendicular to the current and
close to the plane.

The two-photon processes, see Fig. 2b, are suppressed by the factor of $%
\alpha _{QED}=\frac{e^{2}}{c\hbar }\approx \frac{1}{137}$ compared to the
one-photon process due to an additional vertex, while the phase space\ of
two diagrams is of the same order; see Appendix C for details.

\section{Discussion}

Now we elaborate on a number of related issues and comment on the
experimental feasibility of exploring the Schwinger phase physics. We start
with a qualitative discussion of the rather unusual physics arising at
strong applied fields, when the pair recombination becomes an important
relaxation channel.

\subsection{Coulomb interaction and the formation of the neutral
electron-hole plasma}

Even if the ballistic time and the relaxation time are very large,
Schwinger's dynamical pair creation phase cannot persist for a long time at
large field since density of charges (or both signs) becomes large. When the
density of quasiparticles reaches the order of $\rho _{p}=10^{11}cm^{-2}$ a
neutral electron - hole plasma is created \cite{Balev09} (like in some
semiconductor systems under irradiation). In this state electrostatic
interactions (despite being screened at large distances) become dominant, as
was discussed extensively in connection with electron - positron pairs
creation in Quantum Electrodynamics\cite{Ahiezer}. When electrons and holes
are close enough they strongly attract each other effectively facilitating
the recombination process. The rate therefore far exceeds the one calculated
within perturbation theory in Section III. Let us first estimate when this
state is achieved at experimentally accessible situations.

Assuming the Schwinger pair creation rate, Eq.(\ref{Schwinger_rate}), the
density will approach $\rho _{p}$ at times of order $t_{p}\varpropto \rho
_{p}/E^{3/2}$. With a moderate field value of $E=10^{4}V/m$ \cite{SinghPRB09}%
, the 'plasma time' $t_{p}=140ps\simeq 400$ $t_{nl}$ exceeds the ballistic
time of the $L=1\mu m$ long sample (and probably also the relaxation time in
current experiments on graphene). Yet with higher achievable fields $%
E=2\cdot 10^{6}V/m$ \cite{Vandecasteele10}, the plasma time is reduced to $%
t_{p}=40fs\simeq 2$ $t_{nl}\ll $ $t_{bal}$. Therefore one expects that the
"radiation friction" dissipation channel opens up: electron - hole pairs
annihilate emitting photons, which take energy out of the graphene sheet and
thus a new Ohmic behavior is reached. This is roughly the ballistic time
range for which the emission was calculated in Section III. Of course, due
to the Coulomb attraction enhancement of the recombination the intensity
becomes grossly underestimated in the plasma regime. The pair density will
have to be re-calculated via Boltzmann equations; this will be done in a
separate publication. One however might expect qualitatively that this
conductivity\ in the "second" Ohmic regime certainly exceeds $\sigma _{2}$
and is likely to reach several times $\sigma _{2}$. Data presented in ref.%
\cite{Vandecasteele10} for \textit{clean} samples, see Figs. 3 and 12
therein, is consistent with the linear (Ohmic) I-V curve at such
conductivity value. However the experimental situation in the transport
experiment is rather complex as discussed next.

\subsection{Experimental evidence of the pair creation in samples of
mesoscopic size}

In this subsection we use the above "radiation friction" scenario to discuss
whether there is a clear and unambiguous signature of the Schwinger's pair
creation phase in transport experiments. In a series of remarkable
experiments the nonlinear I-V were measured at high electric fields of the
order $2\cdot 10^{6}V/m$ at room temperature\cite{Vandecasteele10}. The
samples were treated in such a way that, despite being non-suspended, the
typical charge asymmetry did not appear and the Dirac point was accessed
convincingly at zero gate voltage (this demonstrates high quality and is in
variance with most samples on substrate). The I-V curves were studied in
various high and low mobility samples (up to $\mu =7000cm^{2}/\left(
Vs\right) $) and effects of disorder were partially controlled by
irradiating the samples. The samples were $L=1-2\mu m$ long and rather
narrow $\left( W=0.5\mu m\right) $ and the four-probe technique was applied.
Although a nonlinear I-V dependence $I\varpropto V^{\alpha }$ with exponent $%
\alpha =1.3-1.5$ was observed at Dirac point, surprisingly the nonlinearity
disappeared in the highest mobility samples. In these experiments the
ballistic time $t_{bal}=2ps$ is much larger than $t_{nl}=20fs$ at the
highest applied voltage of $4V$. Unfortunately the crossover voltages,
namely when $t_{nl}\left( V\right) =t_{bal}$ or 
\begin{equation}
V_{nl}=\frac{\hbar v_{g}}{eL}=0.32mV  \label{exp}
\end{equation}%
were not probed since at room temperature $k_{B}T=25meV$. As argued in the
previous subsection the radiation friction causes a second Ohmic regime and
the I-V curve in the clean samples is expected to be linear.\ It is disorder
that might have caused the observed nonlinearity in irradiated samples. This
requires an additional theoretical study that includes the effect of pair
recombination. As argued above it becomes as important as the
Landau-Zener-Schwinger pair creation process at such currents.

\section{Conclusions}

To summarize, electron - hole pairs are copiously created via
Landau-Zener-Schwinger mechanism near the Dirac points in graphene or
similar 2D electronic systems by an applied electric field, provided the
available ballistic time exceeds $t_{nl}$, Eq.(\ref{t_nl}). The
recombination into photons produces a characteristic signal proportional to $%
E^{3/2}$ at frequencies of order $t_{nl}^{-1}$ that enables unintrusive and
unambiguous experimental observation of the Schwinger phenomenon.\ The
angular and polarization dependence of the emitted photons with respect to
the graphene sheet was calculated. At very high currents and sufficiently
long ballistic times the recombination process becomes greatly enhanced by
the electron - hole attraction and the radiation becomes an effective
channel of dissipation, the radiation friction.

The calculation can be trivially extended to any system with a Dirac point -
like spectrum as double layer graphene and the recently synthesized family
of materials called "topological insulators" \cite{topins} in which surface
excitations are similar to those in graphene with the notable exception of
chirality. Schwinger's mechanism is also expected in these materials since
the mechanism does not involve chirality (left and right movers contribute
equally to the emission rate of graphene). These materials have an advantage
of not being strictly two dimensional, although ballistic times might be
shorter at present. The driving current should not necessarily be DC, a
sufficient condition is $\omega _{ext}<<t_{nl}^{-1}$. Detectors of light
(photon counters) in the microwave-infrared which are sensitive enough have
recently been developed \cite{detector}. Hopefully Schwinger's pair creation
rate formula can be directly tested using novel condensed matter materials
endowed with relativistic fermion spectra.\bigskip

Acknowledgements. We are indebted to E. Farber and W. B. Jian, J. Pan for
valuable discussions. Work of B.R. and H.K. was supported by NSC of R.O.C.
Grants No. 98-2112-M-009-014-MY3 and 98-2112-M-003-002-MY3, respectively,
the National Center for Theoretical Sciences, and MOE ATU program. H.K.
acknowledges the hospitality of the Physics Department of AUCS, while M.L.
acknowledges the hospitality and support of the NCTS.

\section{Appendix A. Phase space of the one-phton process}

Planar electrons and holes are described by their momenta $\mathbf{p}$ and $%
\mathbf{p}^{\prime }$ in the $x-y$ plane, (see Fig.1a and subsection IIA for
notations), while the momentum of the photon $k$ is three dimensional. The
conservation of the in-plane momentum and the conservation of energy read:

\begin{equation}
\mathbf{p}+\mathbf{p}^{\prime }\mathbf{=}\mathbf{k};\text{ \ \ \ \ }%
v_{g}\left( p+p^{\prime }\right) =ck\mathbf{.}  \label{A1}
\end{equation}%
The momentum in the $z$ direction is not conserved; it is balanced by the
elasticity of the graphene flake. Since $k=\sqrt{k_{z}^{2}+\left\vert 
\mathbf{k}\right\vert ^{2}}\geq \left\vert \mathbf{k}\right\vert ,$ in terms
of $\mathbf{p}$ and $\mathbf{p}^{\prime }$ one has the inequality

\begin{equation}
v^{2}(p+p^{\prime })^{2}-\left\vert \mathbf{p}+\mathbf{p}^{\prime
}\right\vert ^{2}\geq 0,  \label{A2}
\end{equation}%
or, in the polar coordinates,%
\begin{equation}
\left( 1-v^{2}\right) \left( p^{2}+p^{\prime }{}^{2}\right) +2\left[ \cos
\left( \phi -\phi ^{\prime }\right) -v^{2}\right] pp^{\prime }\leq 0.
\label{A2a}
\end{equation}%
With the above constraint the condition that $p^{\prime }$ has real
solutions leads to%
\begin{equation}
\cos \left( \phi -\phi ^{\prime }\right) \leq -1+2v^{2}\text{\textbf{.}}
\label{A3}
\end{equation}%
As $v\ll 1,$ it is obvious that $\phi ^{\prime }-\phi $ is very close to $%
\pi .$ By defining $\Delta \phi =\pi -\left( \phi ^{\prime }-\phi \right) ,$
one see the above condition simplifies to

\begin{equation}
-2v\leq \Delta \phi \leq 2v.  \label{A4}
\end{equation}%
Substituting the above result back into Eq.(\ref{A2a}) it can be seen that $%
p^{\prime }$ is very close to $p.$ By introducing $\Delta r=1-p^{\prime }/p$%
, the condition becomes%
\begin{equation}
\Delta r^{2}+\Delta \phi ^{2}\leq 4v^{2}.  \label{A5}
\end{equation}%
Therefore the allowed region is a disk of radius $2v.$

\bigskip

\section{Appendix B. Derivation of the amplitude and spectral emittance}

The Golden rule photon emission rate (for an "initial" electron with
momentum $\mathbf{p}$,\ a "final" hole $\mathbf{p}^{\prime }$ and a photon
of polarization $\lambda $ and momentum $\left( \mathbf{k},k_{z}\right) $)
is given by Eq.(\ref{W_lambda}). Correspondingly the rate defined in Eq.(\ref%
{Idef1}) is

\begin{eqnarray}
&&\mathcal{M}^{\left( \lambda \right) }\left( \theta ,\varphi ,\frac{\omega 
}{c},t\right)  \notag \\
&=&\frac{e^{2}v_{g}^{2}}{\left( 2\pi \right) ^{4}}\int d\mathbf{p}\left\vert 
\mathcal{F}_{\mathbf{p},\mathbf{k}-\mathbf{p}}^{\left( \lambda \right)
}\right\vert ^{2}N_{\mathbf{p}}N_{\mathbf{k}-\mathbf{p}}\delta \left(
v_{g}\left( p+\left\vert \mathbf{k-p}\right\vert \right) -\omega \right) 
\notag \\
&=&\frac{e^{2}v_{g}^{2}}{\left( 2\pi \right) ^{4}}\int p\left( \phi \right)
d\phi \left\vert \mathcal{F}_{\mathbf{p},\mathbf{k}-\mathbf{p}}^{\left(
\lambda \right) }\right\vert ^{2}N_{\mathbf{p}}N_{\mathbf{k}-\mathbf{p}%
}\cdot J\left( \phi \right) .  \label{B1}
\end{eqnarray}%
$\ \ \ \ \ \ \ \ \ \ \ \ \ \ \ \ \ \ \ \ \ \ \ \ \ \ \ \ \ \ \ \ \ \ \ \ \ \
\ \ \ \ \ \ \ \ \ \ \ \ \ \ \ \ \ \ \ \ \ \ \ \ \ $

Because of the delta function, the integration over $\mathbf{p}$ imposes the
condition of energy conservation.\ As a result, the squares of the matrix
elements, Eq.(\ref{matrix_elements}) simplify:

\begin{eqnarray}
\left\vert \mathcal{F}_{\mathbf{p},\mathbf{p}^{\prime }}^{\left( 1\right)
}\right\vert ^{2} &=&\frac{4\left[ \cos \left( \phi -\varphi \right) -v\sin
\theta \right] ^{2}}{\left( 1-2v\sin \theta \cos \left( \phi -\varphi
\right) +v^{2}\sin ^{2}\theta \right) };  \notag \\
\left\vert \mathcal{F}_{\mathbf{p},\mathbf{p}^{\prime }}^{\left( 2\right)
}\right\vert ^{2} &=&\frac{4\cos ^{2}\theta \sin ^{2}\left( \phi -\varphi
\right) }{\left( 1-2v\sin \theta \cos \left( \phi -\varphi \right)
+v^{2}\sin ^{2}\theta \right) }\text{.}  \label{B3}
\end{eqnarray}%
Moreover, there is an additional factor from the delta function: 
\begin{equation}
J\left( \phi \right) =\frac{\left( 1-2v\sin \theta \cos \left( \phi -\varphi
\right) +v^{2}\sin ^{2}\theta \right) }{2v_{g}\left( 1-2v\sin \theta \cos
\left( \phi -\varphi \right) \right) ^{2}}.  \label{B4}
\end{equation}%
The Jacobian of the transition to polar coordinates is%
\begin{equation}
p\left( \phi \right) =\frac{\omega \left( 1-v^{2}\sin ^{2}\theta \right) }{%
2v_{g}\left( 1-2v\sin \theta \cos \left( \phi -\varphi \right) \right) }.
\label{B5}
\end{equation}%
In view of the step functions for the LZS density, Eq.(\ref{LZ}), the
following possibilities occur:

\textit{a. into the forward direction (positive projection on the electric
field) }$0<\varphi $\textit{\ }$<\pi :$

The\ conditions imposed by the step functions are $0<-p_{y}$ and $0<\frac{eE%
}{\hbar }t+p_{y}-k_{y}$. In terms of the polar coordinates in the momentum
space

(i) for $0<\frac{2t}{\omega t_{nl}^{2}}<1+v\sin \theta \sin \varphi ,$ the
allowed regions are $\Delta _{+}-\phi _{0}<\phi \,<0$ or $-\pi <\phi <-\pi
+\phi _{0}-\Delta _{+}$,\ with 
\begin{eqnarray}
\phi _{0} &=&\arcsin \left( \frac{2t}{\omega t_{nl}^{2}}\right) ,\text{ } 
\notag \\
\text{\ }\Delta _{+} &=&v\sin \theta \tan \phi _{0}\left( \cos \left(
\varphi +\phi _{0}\right) +2\sin \varphi \csc \phi _{0}\right) .  \label{B6}
\end{eqnarray}

(ii) for $1+v\sin \theta \sin \varphi <\frac{2t}{\omega t_{nl}^{2}},$ the
allowed region for $\phi $ is $-\pi <\phi \,<0.$

\textit{b. into the backward direction (negative projection on the electric
field) }$-\pi <\varphi $\textit{\ }$<0$:

The\ conditions imposed by the step functions now become $0<-p_{y}+k_{y}$
and $0<\frac{eE}{\hbar }t+p_{y}$. In terms of the polar coordinates in the
momentum space

(i) for $-2v\sin \theta \sin \varphi <\frac{2t}{\omega t_{nl}^{2}}<1-v\sin
\theta \sin \varphi ,$ the allowed regions are $-\left( \phi _{0}-\Delta
_{-}\right) <\phi \,<2v\sin \theta \sin \varphi $ or $-\pi -2v\sin \theta
\sin \varphi <\phi <-\pi +\phi _{0}-\Delta _{-}$, where$\ \ $%
\begin{equation}
\Delta _{-}=v\sin \theta \tan \phi _{0}\cos \left( \varphi +\phi _{0}\right)
.  \label{B6a}
\end{equation}

(ii) for $1-v\sin \theta \sin \varphi <\frac{2t}{\omega t_{nl}^{2}},$ the
allowed region for $\phi $ is $-\pi -2v\sin \theta \sin \varphi <\phi
\,<-2v\sin \theta \sin \varphi .$

Since $v\equiv v_{g}/c\simeq 1/300<<1$, one can neglect higher order
correction in $v$. The conditions for case (i) simplify into $0<\frac{2t}{%
\omega t_{nl}^{2}}<1$ and $-\phi _{0}<\phi \,<0$ or $-\pi <\phi <-\pi +\phi
_{0}.$ Therefore 
\begin{eqnarray}
&&\mathcal{M}^{\left( 1\right) }\left( \theta ,\varphi ,\frac{\omega }{c}%
,t\right) =\frac{e^{2}\omega }{8\pi ^{4}}  \label{B7} \\
&&\times \int_{0}^{\phi _{0}}d\phi \left( \cos ^{2}\phi \cos ^{2}\varphi
+\sin ^{2}\phi \sin ^{2}\varphi \right) \exp \left[ -\frac{\pi }{2}\omega
^{2}t_{nl}^{2}\cos ^{2}\phi \right] \text{;}  \notag \\
&&\mathcal{M}^{\left( 2\right) }\left( \theta ,\varphi ,\frac{\omega }{c}%
,t\right) =\frac{e^{2}\cos ^{2}\theta }{8\pi ^{4}}  \notag \\
&&\times \int_{0}^{\phi _{0}}d\phi \left( \cos ^{2}\phi \sin ^{2}\varphi
+\sin ^{2}\phi \cos ^{2}\varphi \right) \exp \left[ -\frac{\pi }{2}\omega
^{2}t_{nl}^{2}\cos ^{2}\phi \right] \text{.}  \notag
\end{eqnarray}%
Similarly, the conditions for case (ii) simplify: $\frac{2t}{\omega
t_{nl}^{2}}>1$ and$\ -\pi <\phi \,<0.$ Thus

\begin{eqnarray}
&&\mathcal{M}^{\left( 1\right) }\left( \theta ,\varphi ,\frac{\omega }{c}%
,t\right)  \label{B8} \\
&=&\frac{e^{2}}{16\pi ^{4}}\omega \int_{-\pi }^{0}d\phi \cos ^{2}\left( \phi
-\varphi \right) \exp \left[ -\frac{\pi }{2}t_{nl}^{2}\omega ^{2}\cos
^{2}\phi \right]  \notag \\
&=&\frac{e^{2}}{32\pi ^{3}}\omega e^{-\pi t_{nl}^{2}\omega ^{2}/4}\left[
I_{0}\left( \frac{\pi t_{nl}^{2}\omega ^{2}}{4}\right) -\cos 2\varphi
I_{1}\left( \frac{\pi t_{nl}^{2}\omega ^{2}}{4}\right) \right] ;  \notag \\
&&\mathcal{M}^{\left( 2\right) }\left( \theta ,\varphi ,\frac{\omega }{c}%
,t\right)  \notag \\
&=&\frac{e^{2}}{16\pi ^{4}}\omega \cos ^{2}\theta \int_{-\pi }^{0}d\phi \sin
^{2}\left( \phi -\varphi \right) \exp \left[ -\frac{\pi }{2}t_{nl}^{2}\omega
^{2}\cos ^{2}\phi \right]  \notag \\
&=&\frac{e^{2}}{32\pi ^{3}}\omega e^{-\pi t_{nl}^{2}\omega ^{2}/4}\cos
^{2}\theta \left[ I_{0}\left( \frac{\pi t_{nl}^{2}\omega ^{2}}{4}\right)
+\cos 2\varphi I_{1}\left( \frac{\pi t_{nl}^{2}\omega ^{2}}{4}\right) \right]
.  \notag
\end{eqnarray}%
These expressions lead to the final results for the spectral emittance Eq.(%
\ref{I_perp1}) and Eq.(\ref{Iasympt}).

Next we consider the angular and polarization dependence of the radiated
power per unit area defined as the spectral intensity $\mathcal{L}^{\left(
\lambda \right) }\left( \theta ,\varphi ,t\right) $, Eq.(\ref{I_lambda_def}%
). In the above formulae one can first integrate over $\omega $ in the range 
$0<\omega \,<\frac{2t}{t_{nl}^{2}}\csc \phi $, and then integrate over $\phi 
$ in the region $\left[ 0,\frac{\pi }{2}\right] $. This leads to Eqs.(\ref%
{angle1}) and (\ref{angle21a}) for the luminosity integrated over
frequencies.

\bigskip

\section{Appendix C. Phase space of the two-photon process\ }

As in QED \cite{Ahiezer}, the two-photon diagram, Fig. 1b, gives rise to the
following S - matrix element

\begin{eqnarray}
&&F_{2}^{\left( \lambda ,\lambda ^{\prime }\right) }\left( p,p^{\prime
},k,k^{\prime }\right)  \label{C1a} \\
&=&\frac{ie^{2}}{4\sqrt{\omega \omega ^{\prime }\varepsilon \varepsilon
^{\prime }}}\mathcal{F}_{2}^{\left( \lambda ,\lambda ^{\prime }\right) }\
\left( 2\pi \right) ^{4}\delta \left( v_{g}\left( p+p^{\prime }\right)
-c\left( k+k^{\prime }\right) \right) \mathbf{,}  \notag
\end{eqnarray}%
where 
\begin{equation}
\mathcal{F}_{2}^{\left( \lambda ,\lambda ^{\prime }\right) }=v^{\dagger
}\left( -\mathbf{p}^{\prime }\right) \Xi ^{\left( \lambda ,\lambda ^{\prime
}\right) }u\left( \mathbf{p}\right)  \label{C2}
\end{equation}%
and%
\begin{equation*}
\Xi ^{\left( \lambda ,\lambda ^{\prime }\right) }=%
\begin{array}{c}
\left( v\mathbf{\sigma \cdot e}^{\prime \left( \lambda ^{\prime }\right)
}\right) i\frac{\left( k-p^{\prime }\right) -\mathbf{\sigma \cdot }\left( 
\mathbf{k}-\mathbf{p}^{\prime }\right) }{v\left[ \left( k-p^{\prime }\right)
^{2}-\left( \mathbf{k}-\mathbf{p}^{\prime }\right) ^{2}\right] }\left( v%
\mathbf{\sigma \cdot e}^{\left( \lambda \right) }\right) \\ 
+\left( v\mathbf{\sigma \cdot e}^{\left( \lambda \right) }\right) i\frac{%
\left( p-k\right) -\mathbf{\sigma \cdot }\left( \mathbf{p}-\mathbf{k}\right) 
}{v\left[ \left( p-k\right) ^{2}-\left( \mathbf{p}-\mathbf{k}\right) ^{2}%
\right] }\left( v\mathbf{\sigma \cdot e}^{\prime \left( \lambda ^{\prime
}\right) }\right)%
\end{array}%
\end{equation*}

The outcome of the integral in Eq.(\ref{Idef1}) is dictated by the size of
the phase space and by the powers of the small parameters $v=v_{g}/c$ and $%
\alpha _{QED}=e^{2}/\left( \hbar c\right) $. The power of $v$ is the same as
in the one-photon case, see Eq.(\ref{M}). Yet there appears an additional
power of $\alpha _{QED}.$

Let us estimate the phase space. The conservation of momentum and energy for
the two-photon process imply:

\begin{equation}
\mathbf{p}+\mathbf{p}^{\prime }\mathbf{=k+k}^{\prime };\text{ \ \ }%
v_{g}\left( p+p^{\prime }\right) =c\left( k+k^{\prime }\right) \mathbf{.}
\label{C3}
\end{equation}%
Since $\left( k+k^{\prime }\right) ^{2}\geq \left( k_{x}+k_{x}^{\prime
}\right) ^{2}+\left( k_{y}+k_{y}^{\prime }\right) ^{2}$ one still has the
inequality $v^{2}(p+p^{\prime })^{2}-\left\vert \mathbf{p}+\mathbf{p}%
^{\prime }\right\vert ^{2}\geq 0$. Just like in the one-photon case
(Appendix A), utilizing $v<<1$ leads to the constraints $\phi -\phi ^{\prime
}\approx \pi $ and $p^{\prime }\approx p$. The phase space of the two-photon
case is thus roughly of the same order as that of the one-photon process.
Therefore this two-photon process can be neglected.

\end{document}